\documentclass[prd,aps,preprint,eqsecnum,floats,showpacs]{revtex4}
\begin{document}
\title{Leptonic widths of high excitations in heavy quarkonia}
\author{A.M. Badalian and A.I. Veselov}
\affiliation{Institute of Theoretical and Experimental Physics,
B.Cheremushkinskaya 25, 117218 Moscow, Russia, }
\author{B.L.G. Bakker}
\affiliation{Vrije Universiteit, Department of Physics,
Amsterdam, The Netherlands}
\newcommand{\beq}{\begin{eqnarray}}
\newcommand{\eeq}{\end{eqnarray}}
\newcommand{\be}{\begin{equation}}
\newcommand{\ee}{\end{equation}}
\def\la{\mathrel{\mathpalette\fun <}}
\def\ga{\mathrel{\mathpalette\fun >}}
\def\fun#1#2{\lower3.6pt\vbox{\baselineskip0pt\lineskip.9pt
\ialign{$\mathsurround=0pt#1\hfil ##\hfil$\crcr#2\crcr\sim\crcr}}}
\newcommand{\vep}{\mbox{\boldmath${\rm p}$}}
\newcommand{\lan}{\langle}
\newcommand{\ran}{\rangle}
\begin{abstract}
Agreement with the measured electronic widths of the
$\psi(4040)$, $\psi(4415)$, and $\Upsilon (11019)$ resonances is shown
to be reached if two effects are taken into account: a flattening of
the confining potential at large distances and a total screening of the
gluon-exchange interaction at  $r\ga 1.2$  fm. The leptonic widths of
the unobserved $\Upsilon(7S)$ and $\psi(5S)$ resonances:
$\Gamma_{e^+e^-}(\Upsilon (7S))=0.11$ keV and $\Gamma(\psi(5S))\approx
0.54$ keV are predicted.

\end{abstract}
\maketitle

\section{Introduction}

Recently, new experimental data on leptonic widths in heavy quarkonia
(HQ) has been presented \cite{ref.1,ref.2,ref.3}. In the BaBar
experiment the mass and the total and electronic widths of the $\Upsilon
(10580)$ resonance have been measured with great accuracy \cite{ref.1},
while the CLEO Collaboration has observed significantly larger muonic
branching ratios of the $\Upsilon (nS)$ resonances $(n=1,2,3)$
compared to the values adopted till now \cite{ref.4}.  Besides, existing
experimental data on the total cross section for hadron production in
$e^+e^-$ annihilation (in   the region $\sqrt{s} = 3.8 \div 4.8$ GeV)
have been reanalysed  \cite{ref.3}, and the total and electronic widths of
the $\psi(4040), \psi(4160), \psi(4415)$ resonances are shown to be
larger by 20\% for the $\psi(4040)$ and by 70\% for the  $\psi(4415)$
resonance than their values from the Particle Data Group (PDG) \cite{ref.4}.

Accurate knowledge of the leptonic widths of high meson
excitations is of special importance for the theory, because the
wave functions (w.f.) at the origin  of vector $nS$
resonances, $|R_n(0)|^2$, proportional to $\Gamma_{e^+e^-}$,
directly provides information about the static $Q\bar Q$
interaction at all distances including large $r$. In HQ they can
be expressed through the matrix element (m.e) of the static
force. Unfortunately, the true behavior of the  static potential
$V(r)$ at $r\ga 1$~fm is still undefined in QCD and from lattice
measurements there is only an indication that the confining linear
potential $\sigma_0 r$ is becoming more flat at large  $r$
\cite{ref.5}.

The origin of this phenomenon has been discussed in \cite{ref.6} where
it was shown that flattening occurs due to the creation of virtual
$q\bar q$ pairs in the Wilson loop before the string breaking takes
place \cite{ref.6}. Due to the presence of virtual loop(s) the surface
of $\lan W(C)\ran$, and therefore the effective string tension, is
becoming smaller: the string tension $\sigma(r)$  depends on $r$ and
its derivative $\sigma'(r)<0$. For light mesons this phenomenon gives
rise to a correlated shift down of radial excitations which increases
with $n$ (for $\rho(3S)$ this shift is about 150 MeV \cite{ref.6}).
Our chereent alculations show that a similar {\em correlated} shift
down takes place for high excitations in HQ, being about 60 MeV for
$\psi(4415)$ and about 30 MeV for $\Upsilon (6S)$ \cite{ref.7}, while
the leptonic widths of high excitations provide an additional
opportunity to test the confining potential at large $r$.

In this paper we concentrate on the leptonic widths of HQ. For
low-lying resonances they have been calculated in many papers [8-10],
where it has been observed  that agreement with the experimental values
of $\Gamma_{e^+e^-}$  can be obtained only if the asymptotic freedom
(AF) behaviour of the vector coupling $\alpha_V(r)$ in the
gluon-exchange (GE) interaction is taken into account (this effect  is
about 50\%). For the Coulomb interaction $(\alpha_V=const)$ the
leptonic widths of both low- and high-lying resonances in the
$\Upsilon$- and $\psi$-families appear to be 50-100\% higher than
their experimental values.

However, even if the AF behavior of $\alpha_V(r)$ is taken into account
and for  low-lying resonances (like $J/\psi, \psi(2S), \Upsilon(nS)
(n\leq 3)$ the leptonic widths are in agreement with experiment, still
for  very high  excitations (like the $\psi(4040),$ $ \psi(4415),$ $
\Upsilon(11019)$) the calculated $\Gamma_{e^+e^-}$ appear to be
significantly larger than the experimental values. The characteristic
feature of these resonances is that they have very large sizes (their
r.m.s. radii $r_n\ga 1.2$ fm) and therefore their w.f. at the origin
are very sensitive to details of the $Q\bar Q$ interaction at all
distances.  It does not seem accidental that better agreement with
experiment for high resonances is obtained in \cite{ref.11} where a
mild (logarithmic) confining potential (instead of linear $\sigma_0 r$
potential) has been used.

In this paper we study two effects which give rise to a decrease of the
leptonic widths of high excitations in HQ. The first one is the
flattening of the confining potential at large $r$. The second effect
occurs if the GE interaction is very much suppressed or even switched
off due to a screening at distances $r\ga 1.0$ fm. The reason of such a
total screening needs a special analysis \cite{ref.7}, but  the
dynamics of a resonance with large radius, defined  by the confining
potential only, appears to be rather simple.

\section{Leptonic widths as probes of the gluon exchange interaction}

The electronic width of the vector meson $V(nS)$ is given by the
Van Royen--Weisskopf formula \cite{ref.12} with the QCD correction
taken into account \cite{ref.13}. It contains the w.f. at the origin
and some known quantities:
\be
 \Gamma_{e^+e^-} (V(nS)) = \frac{4e^2_Q\alpha^2}{M_n^2(V)}
 |R_n(0)|^2\left(1-\frac{16}{3\pi} \alpha_s (2m_Q)\right),
\label{eq.1}
\ee 
Here for $\alpha_s(2m_c)$ and $\alpha_s(2m_b)$ we use the conventional
values:  $\alpha_s(2m_c)=0.253, \,\alpha_s(2m_b)=0.177$ (e.g. see
\cite{ref.10}).  The w.f. at the origin is proportional to the leptonic
width and on the other hand it can be expressed through the m.e. over
the static force $F(r)=\frac{dV}{dr}$.

In the nonrelativistic (NR) approximation the relation is \cite{ref.14}
\be 
 |R_n^{NR} (0)|^2= m_Q\lan F(r)\ran_{nS}.
\label{eq.2}
\ee  
Here for $m_Q$ the heavy quark pole mass entering NR Hamiltonian must
be used \cite{ref.15}. For relativistic kinematics and a relativistic
Hamiltonian  with the use of the ``einbein approximation''  for the
spinless Salpeter equation instead of Eq.~(\ref{eq.2}) the following
relation can be obtained \cite{ref.16}:
\be 
 |R_n(0)|^2=\omega_Q\lan F(r)\ran_{nS}, 
\label{eq.3}
\ee 
where 
\be 
 \omega_Q(nS) = \lan \sqrt{\vep^2+m^2_Q}\ran_{nS}
\label{eq.4}
\ee 
is the average kinetic energy of a heavy quark, or the quark
constituent mass. For $c$ and $b$ quarks in HQ the difference between
$\omega_Q$ and the pole mass $m_Q$ is about 200 MeV for low-lying
states and about $250\div 300$ MeV for high excitations and this
difference gives about 20\% (5\%) corrections to $|R_n(0)|^2$ in
charmonium (bottomonium).

In the general case the static potential can be presented in the form
\be
 V(r) =r\sigma(r) -\frac43 \frac{\alpha_V(r)}{r} f_{\rm scr} (r).
\label{eq.5}
\ee 
To describe low-lying states (below the open-flavor threshold) it is
sufficient to take a linear confining potential with
$\sigma(r)=const=\sigma_0$ and to put the screening function
$f_{sc}(r)=1.$ For high-lying resonances both  effects--the flattening
of the confining potential and the screening of GE interaction--are
becoming important. We shall consider the effects coming from screening
in detail in our next paper \cite{ref.7}, while here we take the
screening function.
\be 
 f_{\rm scr} = \left\{\begin{array}{ll}
 1,&r<R_{\rm scr},\\
 f_0\exp(-(\sqrt{\sigma }\,r)^{4/3}),& r\geq R_{\rm scr}.\end{array}
 \right.
\label{eq.6a}
\ee
The choice of this function with $R_{\rm scr}\approx 0.6$ fm is
motivated by the analysis of the screening effects in \cite{ref.17}.
Here we take a larger value for the screenig radius: $R_{\rm scr}
\approx 1.0$ fm.

Then  for low and high excitations   one can use different
approximations in Eq.~(\ref{eq.3}). For low excitations  $\sigma'(r)$ is
negligible, $\lan \sigma (r) \ran\approx\sigma_0$, but the
contribution from the derivative  $\alpha'_V(r)$ in Eq.~(\ref{eq.3}) is
important (it reflects  the influence of the AF  behavior of the
coupling $\alpha_V(r)$) and one obtains 
\be 
 |R_n(0)|^2= \omega_Q(n) \sigma_0 +\frac43 \omega_Q(n)
 \left\{ \lan r^{-2} \alpha_V (r)\ran_{nS} -
 \lan r^{-1} \alpha'_V(r) \ran_{nS}\right\}\quad ({\rm small~} n). 
\label{eq.6}
\ee 
For high-lying excitations, on the contrary, the derivative
$\alpha'_V(r)$ is small and, moreover, in bottomonium the negative term
$\lan r \sigma'(r) \ran_{n}$ also remains much smaller than $\lan
\sigma (r)\ran$.  For such resonances effects from the screening of GE
interaction is becoming important:
\beq 
 |R_n(0)|^2 & = &
 \omega_Q(n) \left\{\lan \sigma(r)\ran_{nS}-\lan r\sigma'(r)\ran_{nS} \right.
 \nonumber \\
 & & \quad\quad 
 + \left. \frac43 \lan r^{-2} \alpha_V(r) f_{\rm scr}(r)\ran_{nS}
 -\frac43 \lan r^{-1} \alpha_V(r) f'_{s\rm cr}(r)\ran_{nS}\right\}. 
\label{eq.7}
\eeq
In Tables \ref{Table.1} and \ref{Table.2} we present the HQ leptonic
widths calculated for three potentials still neglecting the screening
effects:

1. For the Cornell potential with the parameters taken from \cite{ref.8}
($\alpha_V(r)=const=0.52$) the leptonic widths are very large,  being
50$\div$ 70\% larger for all $\Upsilon (nS)$ resonances $(n\leq 6)$
than the experimental values.

2. For the potential taken from \cite{ref.15},
\be
  V_B(r)=\sigma_0 r-\frac43 \frac{\alpha_B(r)}{r}, \label{eq.8}
\ee
the vector coupling $\alpha_B(r)$ is defined  in background
perturbation theory. This vector coupling $\alpha_B(r)$ has the correct
perturbative limit at small distances and also possesses the property
of  freezing at large $r$. As seen from Tables 1 and 2 this potential gives
a good description of the electronic widths for many HQ states:
$J/\psi, \psi(3686)$ in charmonium and for all $\Upsilon(nS) $
resonances with exception of the $\Upsilon (6S)$ resonance (the mass
$M_{\exp}(6S) =11019$ MeV).

3. To demonstrate the sensitivity of the leptonic widths to the
behavior of the confining potential at large distances we consider the
``modified'' potential, 

\be V_M(r) =r\sigma (r) -\frac43
\frac{\alpha_B(r)}{r},
\label{eq.9}
\ee  
where the flattening effect is taken into account and  the string
tension $\sigma(r)$  is taken as for light mesons \cite{ref.6} while
the vector coupling $\alpha_B(r)$ is  the same as in the potential
Eq.~(\ref{eq.8}).

\begin{table}[ht]
\caption{\label{Table.1} The leptonic widths (in keV) of the $\Upsilon
(nS)$ resonances for the Cornell potential and the potentials given by
Eqs.~(\ref{eq.8}) and (\ref{eq.9}).}   

\begin{tabular}{|c|c|c|c|c|c|c|}
\hline
Potential  &1S & 2S & 3S & 4S & 5S & 6S \\ 
\hline 
$\sigma_0r-\frac{\kappa}{r}$ $^{a)}$ & 2.60 & 0.94& 0.66 & 0.54 &0.47 & 0.42 \\
\hline 
$ \sigma _0r-\frac43 \frac{\alpha_B(r)}{r} ~^{b)}$& 1.21 & 0.56 & 0.41 & 0.34 & 0.30 & 0.27 \\ 
\hline
$\sigma (r) r-\frac43 \frac{\alpha_B(r)}{r} ~^{c)}$ & 1.14 & 0.54 & 0.40 & 0.32 & 0.27 & 0.24 \\
\hline
experiment \cite{ref.4} &
 1.32(7) & 0.52(8) & 0.48(11) & 0.248(31) & 0.31(7) & 0.130(30) \\
\quad\quad \cite{ref.2} & & & & 0.321(46) &  &  \\
\hline
\end{tabular}

$^{a)}$ From \cite{ref.8} where $\sigma_0=0.1826$ GeV$^2$,
$\kappa=0.52$; $m_b=5.17$ GeV is in fact the constituent mass
$\omega_b$ Eq.~(\ref{eq.4}). \\
$^{b)}$ Here $\sigma_0=0.18 $ GeV$^2$,
$\alpha_B(r)$ is taken from \cite{ref.15}, where
$\Lambda_{\overline{MS}}(2-{\rm loop}) =242$ MeV $(n_f = 5)$, and the pole mass
$m_b(2-{\rm loop})=4.83$ GeV.\\
$^{c)}$ Here $\sigma(r)=\sigma_0g(r)$ is taken from \cite{ref.6} with
$\sigma_0=0.18$ GeV$^2$, $(g(0)=1)$, $\alpha_B(r)$ is taken  as
in footnote ${}^{b)}$.
\end{table}

\begin{table}[ht]
\caption{\label{Table.2} The leptonic widths (in keV) of the $\psi (nS)$
resonances for the  same  potentials as in Table~\ref{Table.1}.}
\begin{tabular}{|c|c|c|c|c|}
\hline
Potential  &1S & 2S & 3S & 4S  \\
\hline 
$\sigma_0 r-\frac{\kappa}{r}$ $^{a)}$ & 8.18 & 3.68& 2.62 & 2.01 \\
\hline 
$\sigma_0 r-\frac43 \frac{\alpha_B(r)}{r}^{b)}$
 & 5.13 & 2.48 & 1.80 & 1.39 \\
\hline 
$\sigma(r) r-\frac43 \frac{\alpha_B(r)}{r} ~^{c)}$ & 5.10 & 2.42 & 1.70 & 1.18
  \\ 
\hline 
experiment \cite{ref.4}   & 5.26(37) & 2.19(15) & 0.75(15) & 0.47(10) \\
 \quad\quad \cite{ref.3}  & & & 0.89(8) & 0.71(10)   \\ 
\hline

\end{tabular}

$^{a)}$  The parameters of the Cornell potential are the same as
in footnote ${}^{a)}$ in Table \ref{Table.1} and $m_c=1.84$ GeV.\\
$^{b)}$  See footnote ${}^{b)}$ in Table \ref{Table.1}; the pole mass $m_c=1.44$
GeV, $\Lambda_{\overline{MS}} = 260$ MeV $(n_f = 4)$.\\
$^{c)}$  See footnote ${}^{c)}$ in Table \ref{Table.1}; the pole mass $m_c=1.44$
GeV, $\Lambda_{\overline{MS}} = 260$ MeV $(n_f = 4)$.\\
\end{table}

From Tables \ref{Table.1} and \ref{Table.2}  one can see that  for the
modified potential Eq.~(\ref{eq.9}) the leptonic widths of the
$J/\psi,$ $\psi(2S)$ and $\Upsilon(nS) (n\leq 5)$  are practically the
same  as for the linear potential $\sigma_0r$ Eq.~(\ref{eq.8}) while
for the higher resonances $(\psi(4040)$, $\psi(4415)$, and
$\Upsilon(11019))$ they are smaller by only  $\sim 10\%$ and still
exceed $\Gamma_{e^+e^-} (\exp)$. The characteristic feature of these
three resonances is their large sizes (even in single-channel
approximation) $r_3(\psi(3S))\approx 1.2$ fm; $r_4(\psi(4S))\approx
1.4$ fm, $ r_6(\Upsilon(6S)\approx 1.4$ fm. There can be two possible
reasons for a further decrease of their leptonic widths. First, one may
think of the coupling of the considered $Q\bar Q$ resonance to an open
meson-meson channel. Comparison of the experimental data with our
calculations show that the $4S$ state, $\Upsilon (10580)$, has a
hadronic shift of about 50 MeV due to coupling to the $B\bar B$ channel
\cite{ref.7}, nevertheless, the  calculated electronic width (see
Table~\ref{Table.1}) appears to be in very good agreement with the new
precision measurements of $\Gamma_{e^+e^-}(\Upsilon(10580))$
\cite{ref.1}. Also for the $\Upsilon(10865)$, the $5S\; b\bar b$ state,
for which the mass is close to the $B_s^*\bar B_s^*$ threshold,
agreement between the calculated and experimental value of
$\Gamma_{e^+e^-}$ is obtained. So, one can assume that open channels do
not drastically change the leptonic width of a resonance considered and
cannot explain the $\sim 70\%$ difference between the theoretical and
experimental leptonic widths for $\psi(4415)$ and $\Upsilon(11019)$. (A
small mixing of the $Q\bar Q$ and meson-meson channels was also
observed in Lattice QCD (second ref.  \cite{ref.5})).

Therefore we assume here that a significant reduction of the leptonic
widths of the  $\psi(4415)$ and $\Upsilon (11019)$  resonances occurs
due to a change in the  static potential: a screening of the GE
interaction at large $r$ and flattening of the confining potential.

\section{Leptonic widths of highly excited resonances}

If the screening of the GE interaction with $f_{\rm scr}(r)$
Eq.~(\ref{eq.6}) is taken into account, then the w.f. at the origin is
defined by the relation Eq.~(\ref{eq.8}) where the contribution from
the GE term  appears to be small ($<10\%)$ for high excitations,  so
that
\be
 |R_n(0)|^2 =\omega_Q(n)\{\lan\sigma(r)\ran_n -
 \lan r\sigma'(r)\ran_n\}\equiv\omega_Q(n)\bar\sigma_n
\label{eq.10}
\ee
Here we  shall use the function $\sigma (r)$ in the form and with the
parameters suggested in \cite{ref.6}. Its characteristic values are
following:
\beq
 \sigma(r) \approx \sigma_0 \;\; &  {\rm for} & r\la 1 \; {\rm fm},
\nonumber \\
 \sigma(r=1.3 \; {\rm fm})  & \approx  & 0.94\,\sigma_0,
\nonumber \\
 \sigma(r=2.5 \; {\rm fm})  & \approx  & 0.78\,\sigma_0,
\nonumber \\
 \sigma(r\ga 4 \; {\rm fm}~)  & = & 0.6 \,\sigma_0.
\label{eq.9a}
\eeq 
i.e. this string tension is slowly decreasing for larger $Q\bar Q$
separations $r$ and has asymptotic value $\sigma_{\rm asym}=
0.6\sigma_0(\approx 0.11$ GeV$^2$ for $\sigma_0=0.18$ GeV$^2$). Our
flattening confining potential continues to  grow (with a smaller
slope), and it significantly differs from the one suggested in
\cite{ref.18}, where the confining potential is taken as a constant
equal to $R_{SC}\,\sigma (R_{SC})$ for $r\geq R_{SC}\approx 1.6$ fm. One
may notice that with such an assumption the quarks in a meson are not
confined and can be liberated.

For this simple asymptotic potential $V_{\rm asym}({\rm large~}r)
=r\sigma(r)$ the constituent masses $\omega_n(Q\bar Q)$ and  $\lan
\sigma(r)\ran_{nS}$ can be calculated easily from the solutions of the
spinless Salpeter equation \cite{ref.15}. For the $\Upsilon (6S)$ and
$\Upsilon (7S)$, using $(\sigma_0=0.18$ GeV$^2$ and $m_b\approx 4.83$
GeV) the following numbers are obtained,
\beq
 \omega_7(b\bar b) & \approx  & \omega_6(b\bar b) = 5.1~{\rm GeV}, 
\nonumber \\
 \lan \sigma (b\bar b, r )\ran_{6S}& = & 0.171~{\rm GeV}^2
\nonumber \\ 
 \lan \sigma (b\bar b, r )\ran_{7S} & = & 0.167~{\rm GeV}^2.
\label{eq.10a}
\eeq
and the term $\langle r\sigma'(r) \rangle$ is relatively small.  Then
from Eq.~(\ref{eq.10})
\beq
 |R_6(b\bar b, 0)|^2 & = & 0.87 ~{\rm GeV}^3,
\nonumber \\
 |R_7(b\bar b, 0)|^2 & = & 0.85
~{\rm GeV}^3.
\label{eq.11} 
\eeq  
and  from Eq.~(\ref{eq.1}) one obtains
\be
 \Gamma_{e^+e^-}(\Upsilon(11.019))=0.12~{\rm keV}.
\label{eq.14a}
\ee
This value is in good agreement with the experimental value
$\Gamma_{e^+e^-}(\Upsilon(6S))=0.130\pm 0.030$ keV [4]. With the
use of Eq.~(\ref{eq.13}) the electronic width of the still unobserved
$\Upsilon(7S)$ can also be predicted:
\be
 \Gamma_{e^+e^-}(\Upsilon(7S))=0.11~{\rm keV},
\label{eq.15a}
\ee
where the value of the  mass, $M_7=11.25$ GeV  (obtained in
single-channel approximation)  has been used. Note that 
the mass difference $M(7S) - M(6S)$ is not small, about 230 MeV.

In charmonium for better accuracy, the negative correction in
$\overline{\sigma}_n$ Eq.~(\ref{eq.9}) coming from the derivative $\lan
r \sigma'(r)\ran$, is becoming larger and gives a contribution  of
$\sim 15\%$. The value of $\bar \sigma$ for the $\psi (4S)$ is $ \lan
\sigma(r)-r\sigma'(r)\ran_{4S}=0.14$ GeV$^2$ and $\lan
\sigma(r)-r\sigma'(r)\ran_{5S}=0.13$ GeV$^2$ for $\psi(5S)$, while the
constituent masses are: $\omega_4(c\bar c) =1.71$ GeV and
$\omega_5(c\bar c)=1.67$ GeV.  Then from Eq.~(\ref{eq.10}) it follows
that

\be
 |R_4(c\bar c,
0)|^2=0.24 ~{\rm GeV}^3,\\ |R_5(c\bar c, 0)|^2=0.22 ~{\rm
GeV}^3,\label{eq.12} 
\ee  
and correspondingly, the electronic widths are
%
\be
 \Gamma_{e^+e^-} (\psi(4415))=0.66 ~{\rm keV},~~\Gamma_{e^+e^-}
 (\psi(5S))=0.54 ~{\rm keV}.
\label{eq.13}
\ee  
For the $\psi(4415)$ resonance our theoretical prediction in
Eq.~(\ref{eq.13}) agrees very well with that from the analysis of
Seth~\cite{ref.3} ($\Gamma_{e^+e^-}(\psi(4415))_{\exp}=0.71\pm0.10$
keV), while both numbers significantly differ from PDG's 
$\Gamma_{e^+e^-}(\psi(4415))=0.47\pm 0.10$ keV.

Our treatment above was done in single-channel approximation when the
possibility of string breaking is neglected, while the creation of
virtual $q\bar q$ pairs is taken into account through the dependence of
$\sigma (r)$ on $r$. Since  at present there is no fundamental
string-breaking theory in QCD, we neither do know what the probability
of string breaking and is nor do we know the probability of the
existence of very high $Q\bar Q$ excitations. Therefore we do not know
what is the  upper limit, or the admittable size $R_{max}$ of a
high-lying resonance (the $Q\bar Q$ string) above which a resonance
cannot exist.

Still, the resonances $\Upsilon (7S)$ and $\psi(5S)$ do not have large
sizes, the splitting $\Upsilon (7S)-\Upsilon(6S)$ is not small,
($\Delta M\sim 230$ MeV), and therefore one may expect them to exist.
In our calculations $\bar r_7(b\bar b)=1.6$ fm and $\bar r_5 (c\bar
c)=1.8$ fm (in single-channel approximation) and their masses (without
a hadronic shift) are  $M(\Upsilon(7S)) \approx 11.24$ GeV,
$M(\psi(5S))\approx 4.63$ GeV. In Eqs.~(\ref{eq.15a}) and
(\ref{eq.13})  their electronic widths,
$\Gamma_{e^+e^-}(\Upsilon(7S))=0.11~{\rm keV},
\Gamma_{e^+e^-}(\psi(5S))=0.54~{\rm keV}$ are given (see
Table~\ref{Table.3}).

\begin{table}
\caption{\label{Table.3} The    leptonic widths (in keV) of highly excited
states in charmonium and bottomonium for the flattening potential
$\sigma(r) r$,  taken from \cite{ref.6} $(\sigma_0=0.18$ GeV$^2$).}
\begin{tabular}{|c|c|c|c|c|c|}
\hline
& $\psi(4040)$ & $\psi(4415)$ & $\psi(5S)$ & $\Upsilon(11019) $& $\Upsilon(7S)$ \\
&  &  & $M\approx 4630$ MeV &  & $M\approx 11250$ MeV \\
\hline
~this paper~ & 0.94 & 0.66 & 0.54 & 0.12 & 0.11 \\
\hline
Exper.~\cite{ref.4} & 0.75(15) & 0.47(10) &  & 0.13(3) &  \\
\quad\quad\quad\cite{ref.3} &0.89(8)& 0.71(10)& &&\\
\hline
\end{tabular}
\end{table}

\section{Conclusions}

The electronic widths of high-lying resonances in HQ are of special
interest for the theory because they provide important information
about the  QCD confining potential at large distances.

Our calculations, performed with a relativistic Hamiltonian, show
that three effects give rise to a decrease of the electronic
widths of vector mesons:\\

(i) The asymptotic-freedom behavior of the vector coupling, which
determines the GE potential, gives a decrease of the leptonic widths of
about 70\% for the $\Upsilon(nS)$ resonances $(n\leq 6)$ and  about
50\% for the $\psi(nS)$ $(n\leq 4)$ resonances.\\

(ii) The flattening of the confining potential at large distances gives
an additional drop ($\sim 15\%)$ in the leptonic widths of HQ but only
for high excitations: $n\geq 4$ for the $\Upsilon(nS)$ family and
$n\geq 3$ for the $\psi(nS)$ states. In this case  good agreement with
experiment is obtained for all $\Gamma_{e^+e^-}(\Upsilon(nS))(n\leq 5)$
and for $\Gamma_{e^+e^-}(J/\psi)$, $\Gamma_{e^+e^-}(\psi(3686))$.\\

(iii) If for some reason the GE interaction is totally switched off for
resonances of large size, then the leptonic widths of very high
excitations, like $\Upsilon (11019)$, $ \psi(4040)$, and $\psi(4415)$,
strongly decrease and appear to be in good agreement with the experimental
data. These three resonances have large r.m.s. radii, $r_n(Q\bar Q)\ga 1.2$ fm 
and their purely nonperturbative dynamics turns out to be
rather simple. It is essential that  here  the string tension
$\sigma(r)$ is taken just the same as  in the light meson analysis of
radial excitations \cite{ref.6}.\\

(iv) The electronic widths and masses of the  still unobserved
resonances: $\Gamma_{e^+e^-} (\Upsilon(7S))=0.11$~ keV
$(M_7(b\bar{b})\approx 11250)$ and $\Gamma_{e^+e^-}(\psi(5S))\approx 0.54 $
keV $( M_5(c\bar c)\approx 4630)$ are predicted.\\

\vspace{1cm}

\acknowledgments
We thank Yu.A. Simonov for fruitful discussions. This work was partly supported by the PRF Grant for leading scientific 
schools, Nr. 1774.2003.2.

\end{document}